\documentclass[a4paper,fleqn,usenatbib]{mnras}

\usepackage{epsfig}
\usepackage{epsf}
\usepackage{graphicx}
\usepackage[T1]{fontenc}
\usepackage{ae,aecompl}
\usepackage{amsmath}    
\usepackage{amssymb}    

\def \etal   {\hbox{\it et~al.\/}}

\title{Long-Wavelength, Free-Free Spectral Energy Distributions from
Porous Stellar Winds 
}

\author[R. Ignace]{R.~Ignace\thanks{E-mail: ignace@etsu.edu} \\
Department of Physics \& Astronomy, 
East Tennessee State University, Johnson City, TN, 37614, USA
}

\date{Accepted XXX. Received YYY; in original form ZZZ}

\pubyear{2015}

\begin{document}
\label{firstpage}
\pagerange{\pageref{firstpage}--\pageref{lastpage}}
\maketitle

\begin{abstract}

The influence of macroclumps for free-free spectral energy distributions
(SEDs) of ionized winds is considered.  The goal is to emphasize
distinctions between microclumping and macroclumping effects.
Microclumping can alter SED slopes and flux levels if the volume
filling factor of the clumps varies with radius; however, the
modifications are independent of the clump geometry.  To what extent
does macroclumping alter SED slopes and flux levels?  In addressing
the question, two specific types of macroclump geometries are
explored:  shell fragments (``pancake''-shaped) and spherical clumps.
Analytic and semi-analytic results are derived in the limiting case
that clumps never obscure one another.  Numerical calculations based
on a porosity formalism is used when clumps do overlap.  Under the
assumptions of a constant expansion, isothermal, and fixed ionization
wind, the fragment model leads to results that are essentially
identical to the microclumping result.  Mass-loss rate determinations
are not affected by porosity effects for shell fragments.  By
contrast, spherical clumps can lead to a reduction in long-wavelength
fluxes, but the reductions are only significant for extreme volume
filling factors.

\end{abstract}


\section{Introduction} 

The issue of structured winds from massive stars stretches back
several decades, and the prominence of wind ``clumping'' for
influencing observables has steadily grown
\citep[e.g.,][]{2008cihw.conf.....H}.  Evidence for clumping comes
from a variety of indicators.  The favored mechanism for explaining
the winds of massive stars is the line-driven wind theory
\citep{1970ApJ...159..879L,1975ApJ...195..157C, 1986A&A...164...86P}.
Highly supersonic, massive star winds are driven outwards by the
action of the UV-bright luminosities acting on metal line opacities.
But the process is intrinsically unstable and naturally leads to
structured flows
\citep{1970ApJ...159..879L,1980ApJ...241..300L,1988ApJ...335..914O,
1989ApJ...347.1090M,1997A&A...322..878F,2003A&A...406L...1D,
2005A&A...437..657D}.  Observationally, the evidence for structured
massive-star winds is multi-wavelength in nature, including X-rays
\citep[e.g.,][]{1983ApJ...271..681C,2003A&A...403..217F,
2013ApJ...763..143N}, the UV band, in the form discrete/narrow
absorption components
\citep[e.g.,][]{1986ApJS...61..357P,1995ApJ...452L..53M,
2012MNRAS.422.3142P}, and attempts to explain certain UV-line
strengths
\citep[e.g.,][]{2005A&A...438..301B,2006ApJ...637.1025F,2007A&A...476.1331O,
2008ApJ...685L.149Z,2010A&A...510A..11S}, optical recombination
lines \citep{1991A&A...247..455H,1999ApJ...514..909L}, infrared
excesses \citep[e.g.,][]{2009AN....330..717I,2009AJ....138.1003B},
longer wavelength emissions
\citep[e.g.,][]{1981ApJ...250..645A,2006A&A...454..625P,1997A&A...323..886B,
1998A&A...333..956N,2008A&A...477..373G}, and polarimetry
\citep[e.g.,][]{1987AJ.....93..214L,1991AJ....102.1197T,1995A&A...295..725B,
2007A&A...469.1045D,2009RAA.....9..558L}.

The focus of this contribution is to expand the considerations of
how ``macroclumping'' or ``porosity'' can influence observables of
massive star winds in the infrared (IR) and radio regimes.  In
relation to the effect of clumping for inferring mass-loss rates,
$\dot{M}$, many researchers invoke ``microclumping''.  Microclumping
is the limit in which the radiative transfer is not impacted by the
presence of clumping.  For microclumping all clumps must be optically
thin, and the radiative transfer calculations proceed essentially
as if the flow is ``smooth''.  The influence of microclumping emerges
in terms of the volume filling factor of clumps $f_V$, or an emissive
enhancement factor \citep[which is $1/f_V$,
e.g.][]{1998A&A...335.1003H,2003ApJ...596..538I}.  The presence of
clumping leads to an increase of the source flux relative to a
smooth wind, and for microclumping the enhancement is unconnected
to the specifics of the clump geometry, such as whether clumps take
the form of spheres, filaments, flattened ``pancake'' structures,
or any other shape.

By contrast, macroclumping deals with the case when the radiative
transfer is influenced by the geometry of the clump structures
\citep[e.g.,][]{2004A&A...426..323B}.  For example, based on the
expectations of the line-driving instability operating in massive
star winds, \cite{2004A&A...422..675O} invoked ``shell fragments'',
in the shape of flattened pancakes, to model a time-averaged spherical
distribution of wind shocks to explain resolved X-ray emission
profile shapes observed in massive stars.  In a smooth wind, the
line shapes should be significantly asymmetric \citep[e.g.,]
[]{2001ApJ...549L.119I,2001ApJ...559.1108O,
2002ApJ...568..954I,2007ApJ...659..642L}.  Clumping can alter the
line profile shape for the emergent X-ray radiation.  In particular,
it can reduce line asymmetry
\citep[e.g.,][]{2003A&A...403..217F,2004A&A...422..675O,2006ApJ...648..565O}.

In relation to the UV, initial results from a FUSE study of the
P{\sc v} doublet from massive star winds suggested dramatic reductions
in mass-loss rates of O star winds owing to severe levels of clumping
\citep{2006ApJ...637.1025F}. The argument was that mass-loss rates
from line profile fitting of UV lines is independent of wind clumping
for an ion in its dominant stage.  Comparing $\dot{M}$ values from
P{\sc v} with those derived from clumping-dependent diagnostics
(such as radio emissions) provides a measure of the volume filling
factor for microclumping.  However, the downward revisions were so
severe as to change the expectations of stellar evolution models
substantially \citep{2008cihw.conf....9H}.  One resolution\footnote{
Macroclumping is not the only potential resolution.  The presence
of X-ray emissions could invalidate the assumption that P{\sc v}
is indeed the dominant ion stage for that element, as for example
\cite{2010ApJ...711L..30W} and \cite{2012MNRAS.427...84K}.} to the
problem was found in a consideration of macroclumping
\citep{2007A&A...476.1331O,2013A&A...559A.130S}.

This paper extends consideration of macroclumping effects to the
IR and radio bands.  Others have also explored macroclumping for
radio emissions, such as \cite{1997A&A...323..886B} and
\cite{2008A&A...477..373G}.  The former considered the impact of
discrete shells and shell fragments on the long wavelength emission
(but only for single shells or fragments).  The latter considered
an evolving shell, with application toward explaining variable radio
emissions from P~Cygni.  

In this paper two particular clump geometries are constrasted:
shell fragments and spherical clumps.  Spherical clumps loosely
represent the kinds of structures that can evolve from Rayleigh-Taylor
instabilities.  Rayleigh-Taylor instabilities lead to filamentary
structures that can further devolve into ``knots'' and roundish
clumps \citep[e.g.,][]{2012ApJ...755..160E}.  For example, such
effects can have relevance for the interaction of a wind and the
interstellar medium \citep[e.g.,][]{2012A&A...541A...1M}.  The shell
fragments have long been a favorite for the highly supersonic massive
star winds that are thought to be filled with shock structures
\citep[e.g.,][]{2002A&A...381.1015R}, although more recent simulations
and observations now favor spherical clumps
\citep{2003A&A...406L...1D,2013ApJ...770...80L}.  Discussion 
of the problem begins
with a review in \S~\ref{sec:review} of long-wavelength
emissions from free-free opacity in a strictly smooth wind and in one
with microclumping.  an exploration of porosity
effects for the IR/radio band in \S~\ref{sec:macro} for the two
different selected geometries.  Concluding
remarks are given in \S~\ref{sec:conc}.

\section{SEDs from Smooth or Effectively Smooth Winds}
\label{sec:review}

The objective of this work is to explore the consequences of
macroclumping for IR/radio spectral energy distributions (SEDs)
from ionized stellar winds.  To do so, a number of simplifications
are imposed to aid comparison of cases.  For example, the model
SEDs are restricted to free-free opacity.  It is possible to include
bound-free opacity as well, but at long wavelengths, its addition
will have a similar scaling as the free-free opacity and will not
impact qualitative and comparative trends described here.  A focus
on long wavelengths ensures that the wind will be optically thick,
so thick that a pseudo-photosphere forms in the wind.  In this case
the continuum emission formed in the wind dominates the highly
absorbed stellar emission at the wind base.  Although massive star
winds have overall declining temperature and varying ion fractions
\citep[e.g.,][]{1989ApJS...71..267D}, the wind at large radius tends
to be more nearly isothermal with slowly varying ion fractions.
For this study it is convenient to assume that the wind is isothermal
with fixed ionization.  Also, the flow is assumed to expand radially
at constant speed (i.e., the wind terminal speed, $v_\infty$).

Central to modeling the SEDs is the frequency-dependent optical
depth, $\tau\nu$.  The optical depth from a distant observer to a
point in the wind that lies along the line-of-sight (los) to the
star center is given by

\begin{equation}
\tau_\nu = \int_{\tilde{r}}^\infty \, \kappa_\nu \rho\,R_\ast\,d\tilde{r}'.
	\label{eq:deftau}
\end{equation}

\noindent where $\kappa_\nu$ is the absorption coefficient and
prime indicates a ``dummy variable'' for integration purposes.
Tildes signify lengths that are normalized to the stellar
radius, as for example $\tilde{l} = l / R_\ast$.  
The one exception will be the inverse radius, $u=R_\ast/r$,
that will be used in a number of the integral expressions to be
derived.

Working at relatively long wavelengths for hot-star winds, the
Rayleigh-Jeans limit is adopted, for which $h\nu \ll kT$, and

\begin{equation}
\kappa_\nu \rho = 0.018 \frac{Z_{\rm i}^2}{\mu_{\rm i}\,\mu_{\rm e}}\,
	\frac{\rho^2}{m_H^2}\,T^{-3/2}\,g_\nu\,\nu^{-2},
\end{equation}

\noindent where $Z_{\rm i}$ is the rms ion charge, $\mu_{\rm i}$
and $\mu_{\rm e}$ are mean molecular weights per free ion and per
free electron, respectively, $\rho$ is the mass density of the gas,
$m_H$ is the mass of hydrogen, $T$ is the gas temperature, $g_\nu$
is the free-free Gaunt factor, and $\nu$ is the frequency of
observation.  The adopted assumptions imply that $Z_{\rm i}$,
$\mu_{\rm i}$, and $\mu_{\rm e}$ are constants in the flow for this
study.

The radial optical of equation~(\ref{eq:deftau})can be expressed
as

\begin{equation}
\tau_\nu = \tau_0(\lambda)\,\int_{\tilde{r}}^\infty\,[\rho(\tilde{r}')
	/\rho_0]^2\,d\tilde{r}',
	\label{eq:los_tau}
\end{equation}

\noindent where $\rho_0$ is the density at the base of the wind,
and $\tau_0 = \kappa_0(\lambda)\,\rho_0\,R_\ast$ is the characteristic
optical depth scale as a function of wavelength, with $\kappa_0(\lambda)
= \kappa(\lambda)\,\rho(\tilde{r})/\rho_0$.  The latter is given
by

\begin{equation}
\tau_0(\lambda) = 5\times 10^5\, \frac{Z_{\rm i}^2}{\mu_{\rm i}\,
	\mu_{\rm e}}\,
	\rho_0^2\,T_0^{-3/2}\,
 \left(\frac{R_\ast}{10R_\odot}\right)\,g_\nu\,\lambda_{\rm cm}^2,
\end{equation}

\noindent where fiducial density and temperature values of $\rho_0
= 10^{-13}$~g~cm$^{-3}$ and $T_0=10^4$~K have been assumed, the
stellar radius has been expressed in terms of ten solar radii, and
the wavelength is given in centimeters.  For a wind of completely
ionized, pure hydrogen, $\rho_0$ would correspond to a number density
of about $10^{11}$~cm$^{-3}$, which is characteristic of a wind
with a mass-loss rate of $\dot{M} \sim 10^{-6}~M_\odot$~yr$^{-1}$
and terminal speed $10^3$~km~s$^{-1}$, a fairly massive, evolved
wind from an OB supergiant \citep[e.g.,][]{2012A&A...537A..37M}.
For such a wind, which is typical of the more extreme wind cases
like WR stars, some OB supergiants, or LBV stars, the optical depth
at 1~cm is enormous at the level of $10^6$.  

\subsection{Smooth Winds}

Seminal works for the radio emissions from ionized, massive-star
winds include
\cite{1975MNRAS.170...41W} and \cite{1975A&A....39....1P}.
Both considered how to relate the radio emission to the wind
mass-loss rate, assuming a smooth, spherical wind.  The key results
are that for an observer sightline through the spherical wind, the
emergent intensity is

\begin{equation}
I_\nu = B_\nu(T)\,\left[1-e^{-\tau_{\rm tot}(\tilde{p})}\right],
\end{equation}

\noindent where isothermality is assumed, $\tau_{\rm tot}$ is
the total optical depth of the wind along a ray of normalized impact
parameter $\tilde{p}$, and $B_\nu$ is the Planck function.  For a star
of radius $R_\ast$ at distance $D$ from Earth, the emergent
flux of radiation from the wind is then

\begin{equation}
F_\nu = 2\pi\,\frac{R_\ast^2}{D^2}\,B_\nu(T)\,\int_0^\infty\,
	\left[1-e^{-\tau_{\rm tot}(\tilde{p})}\right]\,\tilde{p}\,d\tilde{p}.
\end{equation}

\noindent At long wavelengths where the wind density is an inverse
square law, and the wind emission can be considered to dwarf the
stellar emission (which is also highly absorbed), the optical depth
factor takes on the form $\tau_{\rm tot} = K(\lambda)\,\tilde{p}^{-3}$,
where the stellar, wind, and wavelength parameters of the opacity
are collected in the wavelength-dependent factor $K$.  In this case
the integral can be analytically evaluated.

\citet{1977ApJ...212..488C} described how free-free
flux that forms in the wind could be interpreted in terms of a
pseudo-photosphere.  Their argument was to
evaluate an effective photospheric radius $r_{\rm eff}$ where
$\tau_\nu = 1/3$ along the los.  From equation~(\ref{eq:los_tau}),
one obtains

\begin{equation}
r_{\rm eff} = \left[3\,\tau_0(\lambda)\right]^{1/3}\,R_\ast \propto
	g_\nu^{1/3}\,\lambda^{2/3}\, R_\ast.
\end{equation}

\noindent Using fiducials of $\rho_0 = 10^{-13}$ g cm$^{-3}$, $T_0 =10^4$~K,
and assuming $g_\nu=1$, the effective radius of the pseudo-photosphere
will be $r_{\rm eff} \approx 80 R_\ast$ at a wavelength of 1~cm.

The flux then becomes

\begin{equation}
F_\nu = \frac{L_\nu}{4\pi\,D^2} = \frac{1}{4}\,B_\nu(T)\,r_{\rm eff}^2
	\propto g_\nu^{2/3}\,\lambda^{-2/3}.
\end{equation}

\noindent With $g_\nu \propto \lambda^{0.11}$ in the radio band
\citep{2000asqu.book.....C}, the radio SED is a power law with a
logarithmic slope exponent of about $-0.6$ with $\lambda$.  Observations
of this value generally signal that the wind is isothermal,
spherical, and at terminal speed.  Slight deviations, especially
somewhat steeper negative slopes, may indicate variations in the
temperature or ionization of the wind.  For the Rayleigh-Jeans
limit, \cite{1977ApJ...212..488C} generalized their results to
relate an observed SED slope in terms of power-law exponents for
the density and temperature distributions.  

Alternatively, deviations from the standard value of $-0.6$ may also
indicate that the free-free emission forms in the wind acceleration
zone, whoch is relevant to lower density winds.  A rising velocity
steepens the density with consequence for the observed SED.  A
positive slope would be non-thermal, generally interpreted as related
to synchroton emission and the presence of magnetism in the extended
wind \citep[e.g.,][]{2011BSRSL..80...67B}.

In this paper the wind is assumed isothermal, with an
ionization structure that is constant throughout, and for
simplicity the sphereical wind will be taken to expand
at constant speed.  In this situation the formal solution
for the optical depth along any sightline traversing
the spherical wind is given by

\begin{equation}
\tau(z,\tilde{p}) = \frac{\tau_0(\lambda)}{2\tilde{p}^3}\, \left[
	\theta - \sin (\theta)\,\cos(\theta) \right],
	\label{eq:tsmooth}
\end{equation}

\noindent where $\tilde{z} = \tilde{p}/\tan(\theta)$. 

In the limit of large optical depth with $\tau_0 \gg 1$, the
``pseudo-photosphere'' formed by the wind completely absorbs the
direct emission by the stellar photosphere, and the continuum
emission is considered ``wind-dominated''.  At wavelengths where
$\tau_0 \ll 1$, the wind emission is small, the absorption of direct
starlight is negligible, and the continuum is said to be
``star-dominated''.  Although model SEDs will properly account for
the starlight component, the discussion of fluxes will generally
focus on the wind contribution and ignore the the stellar contribution.

The integration for flux requires the total optical depth, $\tau_{\rm
tot}$ for a los of fixed impact parameter $\tilde{p}$.  Its value
is given by equation~(\ref{eq:tsmooth}) for $\theta=\pi/2$, with
$\tau_{\rm tot} = \pi\,\tau_0/2\tilde{p}^3$.  Ignoring
the stellar contribution as justified above, the solution for the
flux of emission by the wind is

\begin{equation}
F_\nu \approx 2\pi\,\frac{R_\ast^2}{D^2}\,B_\nu(T_{\rm w})\,
\int_0^\infty\,\left(1-e^{-\pi\,\tau_0/2\tilde{p}^3}\right)\,\tilde{p}\,d\tilde{p},
\end{equation}

\noindent Appendix~\ref{appA}) details the steps to obtaining an
analytic solution for this integral, with the result,

\begin{equation}
F_{\rm sm} = F_\nu = \frac{2}{3}\,\Gamma\left(\frac{1}{3}\right)\times \,
	\frac{\pi\,R_\ast^2}{D^2}\,B_\nu(T_{\rm w})\,\left[
	\frac{\pi\tau_0(\lambda)}{2}\right]^{2/3},
	\label{eq:smoothflux}
\end{equation}

\noindent where the subscript ``sm'' signifies the smooth wind
result, and $\Gamma$ is the ``Gamma'' function.  \cite{1975MNRAS.170...41W}
showed that this expression corresponds to a pseudo-photosphere of
effective radius where the line-of-sight optical depth achieves
$\tau(\lambda) = 2\pi^2/81=0.244$.  \cite{1983ApJ...271..221H}
considered a different measure.  They estimated the optical depth
at which a volume integral of the emissivity would achieve the
required flux.  This ``effective volume'' corrresponds to an
integration of the free-free emissivity beyond $\tau(\lambda) =
(2/3\pi)\, (3/8)^{3/2} = 0.049$. 

\subsection{Winds with Microclumping}

A feature of the radio-band emission is the opportunity to infer
the mass-loss rate of the wind.  The scale of the flux 
is set by $(\dot{M}/v_\infty)^{2/3}$.  Since the wind terminal speed
can be independently constrained from wind-broadened lines
\citep[e.g.,][]{1991ApJ...383..466P}, observed radio fluxes can
be used to infer wind $\dot{M}$ values, which has been pursued by
numerous researchers
\citep[e.g.,][]{1986ApJ...303..239A,1989ApJ...340..518B,
1995ApJ...450..289L}.
However, evidence for the influence of clumping effects became
apparent through a variety of diagnostics, as described in \S~1.
The question arises of how clumping impacts the long-wavelength
emissions.  

Many have adopted a model of ``microclumping'' to infer better
$\dot{M}$ values from radio studies of dense winds.  Microclumping
is the limit in which all individual clump structures are optically
thin. The radiative transport through such a flow is the same as
for a smooth wind, but with a correction factor.  Thus microclumping
is ``effectively smooth'' in its treatment of the flux calculation.
Microclumping does not alter the SED slope as compared to a smooth
wind approach \citep[e.g.,][]{1998A&A...333..956N}, but it does
alter inferred values of $\dot{M}$.

Because the free-free opacity scales as $\rho^2$, clumping
generates more optically thin emission per unit volume as compared
to an unclumped wind.  Defining $f_V$ as the volume filling factor
for the clumping, mass-loss rates are lowered by the multiplicative
factor, $\sqrt{f_V}$ (i.e., for $f_V$ constant throughout the
wind), as compared to no clumping.  This is derived
next, and will be used as a reference against which to compare
macroclumping results.

Let $\langle \rho \rangle$ be the volume-averaged wind density.
Let $\rho_{\rm sm}$ be the unclumped, smooth wind density for a
spherical wind with

\begin{equation}
\rho_{\rm sm} = \frac{\dot{M}}{4\pi\,r^2\,v(r)},
\end{equation}

\noindent where $v(r)$ is the velocity law of the wind.
The density of clumps with microclumping becomes

\begin{equation}
\rho_{\rm mic} = \frac{1}{f_V}\times\frac{\dot{M}}{4\pi\,r^2\,v(r)}.
\end{equation}

\noindent The form of the wind speed will be irrelevant, as
the focus of applications is on asymptotic results for
thick winds where the observed free-free emission forms at
large radii.  At large radius, $v(r) \approx v_\infty$.

When calculating the optical depth integral, the
factor $\kappa_\nu \rho \propto \rho_{\rm mic}^2 \propto f_V^{-2}$.
However, the clumps are encountered only over a fraction of
the integration path length, namely the fraction $f_V$.
Consequently, for microclumping, the optical depth becomes

\begin{equation}
\tau_{\rm mic} = \tau_{\rm sm}/f_V,
\end{equation}

\noindent where $\tau_{\rm sm}$ is the optical depth for a smooth wind
of the same $\dot{M}$, $R_\ast$, and wind velocity
profile.  The expression for the flux of emission, when the
wind is optically thick follows from equation~(\ref{eq:smoothflux}):

\begin{equation}
F_{\rm mic} = \frac{1}{f_V^{2/3}} \times F_{\rm sm}
\end{equation}

\noindent which is the well-known result that for a given 
mass-loss rate, microclumping enhances the flux of emission
by a factor $f_V^{-2/3}$.
To ignore wind clumping would imply  overestimating
the wind mass-loss rate by a factor $f_V^{-1/2}$, since
$F_{\rm sm} \propto \dot{M}^{4/3}$ and $F_{\rm mic}
\propto (\dot{M}^2/f_V)^{2/3}$.

However, there is the possibility that the clumps are not optically
thin, in which case the radiative transport must take account of
the clump geometry \citep[e.g.,][]{2004A&A...413..959B}, a situation
that has been dubbed macroclumping \citep[e.g.,][]{2007A&A...476.1331O}.
The opacity at long wavelengths scales as $\lambda^2$.  Clump
structures that are optically thin at shorter wavelengths can become
optically thick at longer ones, thus the degree to which stochastically
structured flows may be treated as microclumping versus macroclumping
is a $\lambda$-dependent consideration.

\section{Winds with Macroclumping:  Special Case with Discrete Clumps}

The radiative transfer through macroclumps requires specification
of the clumps themselves, as for example the investigations of
\cite{2003A&A...403..217F} and
\cite{2004A&A...422..675O}.  Ultimately a realization of the
structured flow must be imposed, and then the radiative transport
can be numerically evaluated by shooting rays through the medium
to compute emergent intensities and then fluxes for unresolved sources.
\cite{2004A&A...422..675O} use a discrete approach for the simulation
of X-ray observables.

An alternative approach has been described under the heading
of porosity.  \cite{2006ApJ...648..565O}
simulate the effects of macroclumping with integral relations
based on probabilistic considerations associated with the radiative
transfer.  The approach relies on constructing an effective
opacity for expressing the consequences of macroclumping
\citep[e.g.,][]{2012MNRAS.420.1553S,2014A&A...568A..59S}.
The porosity formalism is adopted in the radiative
transport for calculations of the free-free fluxes from winds.

However, before pursuing a solution for the full radiation transport
through the clumped outflow, there is a special circumstance that can
be instructive to consider.  Imagine that clumped structures
are sparsely distributed such that from the observer point-of-view,
no clumps overlap each other as projected on the sky.  Further
suppose that the solution for a given clump is $\delta F_{\nu,{\rm
i}}$. Then the flux for the ensemble total becomes

\begin{equation}
F_\nu = \sum_{\rm i}^{\rm N}\,\delta F_{\nu,{\rm i}}.
\end{equation}

\noindent In the limit that clumps are numerous, the sum can
be represented in integral form:

\begin{equation}
F_\nu = \int\, n_{\rm cl}(\tilde{r},\mu,\phi)\,R_\ast^3\,\delta 
	F_\nu(\tilde{r},\mu,\phi)\,\tilde{r}^2\,d\tilde{r}\,d\mu\,d\phi,
	\label{eq:sparseflux}
\end{equation}

\noindent where $\mu = \cos(\theta)$ and $n_{\rm cl}$ is the number
density of clumps distributed throughout the wind.  

The models under consideration will be for large optical depths for
asymptotic conditions with the emission forming at large radius
where the wind is at terminal speed $v_\infty$.  For a uniform
distribution, the number density of clumps is

\begin{equation}
n_{\rm cl} = \frac{\dot{N}_0}{4\pi\,r^2\,v_\infty},
\end{equation}

\noindent where $\dot{N}_0$ is the injection rate of 
clumps into the flow at the wind base.

Note that the complete absence of overlap among the ensemble is
actually too strong a requirement.  What matters is that the
cumulative optical depth arising from any overlap be optically thin.
This was the limit implicitly used in the application by \citet[][hereafter
``IC'']{2004ApJ...610..351I} to explain radio SEDs
from ultracompact H{\sc ii} regions.  This special case will be
referred to as the ``sparse'' limit, even though the overlap of
structures that are optically thin is formally allowed.

In order to employ the sparse limit expressions, the nature of the
macroclumps must be specified.  Here two cases are considered: shell
fragments and spherical clumps.  Both will be taken as having
constant density within their boundaries, but with densities that
depend on location in the wind flow.  A spherical clump offers the
same flux no matter from what direction it is viewed, but the
intensity distribution across the clump is not uniform (i.e., spherical
clumps are brighter along their diameter than at their limb).
By contrast the shell fragments will be treated as circular
``pancake'' shapes.  The projected shape of a shell fragment when viewed
obliquely is elliptical.  In contrast to spherical clumps,
the intensity across a shell fragment is uniform (i.e., ignoring
edge effects given that the fragments are assumed to be geometrically
thin).  Evidently
different clump structures offer different source properties, and
these can impact their ensemble trends.

\subsection{Sparse Flattened Shell Fragments}

Shell fragment clumps, with outward normals directed along radials
from the star, may represent compressed structures from shocked gas.
One-dimensional, time-dependent, hydrodynamic simulations predict
highly structured winds in terms of dense shells separated by zones
of highly rarified gas.

However, the 1D models overpredict the variability of X-rays from
massive stars.  Although results for $\zeta$~Pup clearly indicate
X-ray variability, the number of discretized, non-spherical clumps
implied seems large \citep{2013ApJ...763..143N}.  Two-dimensional
simulations by \cite{2003A&A...406L...1D,2005A&A...437..657D} suggest
that a highly structured flow can form.  Although 3D models have
not been presented, 1D and 2D simulations have led to a picture of
a wind flow in which the majority of the matter exists in a fairly
random distribution of clumps.

\cite{2003A&A...403..217F} described the radiative transport of
X-ray emissions through shell fragments.  Although at IR and radio
wavelengths, the opacity is different ($\rho^2$ for free-free opacity
versus $\rho$ for photoabsorptive opacity of X-rays), the geometrical
considerations are similar.  One important difference is that for the X-ray
problem, the pancake structures consist of a hot plasma component
adjacent to a cool one.  It is the cool one that does the photoabsorbing;
the hot plasma produces optically thin emission.  For the free-free
situation, the emission and absorption both occur in the same
material, in this case the cooler component\footnote{Since the
free-free opacity drops as $T^{-3/2}$, and the X-ray emitting plasma
is two orders of magnitude hotter than the cooler gas, any hot
component existing at large radius as an ``inter-clump'' component
will be optically thin compared to the cool clump component.
Also, the optically thin free-free emissivity scales as density squared.
The hot component can reasonably be ignored as a contributor to
the long-wavelength free-free emission.}.

\begin{figure}
\includegraphics[width=\columnwidth]{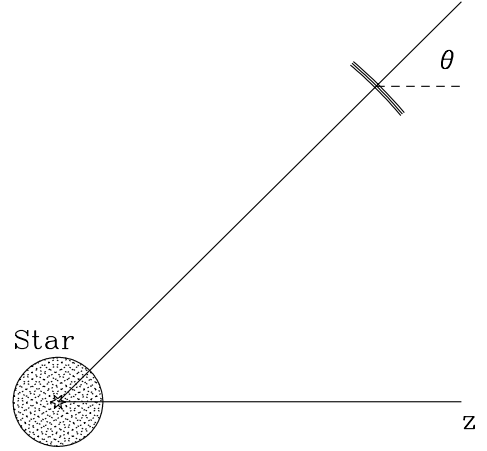}
\caption{Geometry for the shell fragment clump.  In cross-section
a shell fragment is a thin line that is normal to a radial from
the star.  Here the observer is along the $z$-axis and views the
shell fragment at an oblique angle $\theta$.
\label{fig1}}
\end{figure}

Following Feldmeier \etal, the optical depth through a shell fragment
along a ray that intersects it at an oblique angle $\theta$
from the normal will be given by

\begin{equation}
\tau_{\rm z} = \frac{\tau_{\rm r}}{|\mu|},
\end{equation}

\noindent where $\mu = \cos \theta$, $z$ signifies the observer
axis (see Fig.~\ref{fig1}), and $\tau_{\rm r}$ is the optical depth
normal to the shell fragment.  Note that this optical depth is taken
as a constant across the fragment.

For the optical depth along a radial that passes through a shell
fragment, it is tempting to assume that the density is inverse
square by assuming that the fragment evolve through the flow at
constant solid angle, $\delta \Omega$.  However, this is not obvious.
The fact that the fragment represents a shock implies that matter
may accumulate in it, which would make the density less steep than
inverse square \citep[e.g.,][]{2012ASPC..465..140G}.  On the other
hand, the lateral width of the fragment can expand owing to gas
pressure, resulting in an evolving solid angle for the fragment.

For simplicity the density within a fragment is taken to decline
with the radius of its center as a power law, with $\tilde{r}^{-q}$.
The optical depth is

\begin{equation}
\tau_{\rm z} = \tau_0(\lambda)\,\mu^{-1}\,\delta\tilde{r}^{-2q},
\end{equation}

\noindent where 

\begin{equation}
\tau_0 = \kappa_0\,\rho_0\, R_\ast\,\delta \tilde{r}_0,
\end{equation}

\noindent with $\delta r_0$ the radial width of a fragment at
the base of the wind.

The intensity along a ray that is oblique through a fragment is
given by:

\begin{equation}
I_\nu = 2\pi\, B_\nu \, \left[ 1 - e^{-\tau_0\,\mu^{-1}\,\tilde{r}^{-2q}} \right].
\end{equation}

\noindent Again, this assumes that the radial optical depth
$\tau_{\rm r}$ is constant across
the face of the shell fragment.  The flux 
is $\delta F_\nu = \delta \Omega\,\mu\,I_\nu\,(r^2/D^2)$.

Given a time-averaged
spherically symmetric and sparse distribution of clumps, and using
equation~(\ref{eq:sparseflux}), the solution for the
total flux from an ensemble of such structures, all of the same solid
angle extent, becomes

\begin{equation}
F_\nu = B_\nu\, \delta \Omega\,\frac{R_\ast^2}{D^2}\,\frac{\dot{N}_0}{v_\infty
	/R_\ast}\,\int_0^1\,\int_1^\infty \,
	\left[ 1 - e^{-\tau_0\,\mu^{-1}\,\tilde{r}^{-2q}} \right]
	\, \tilde{r}^2\, d\tilde{r}. 
	\label{eq:nocovfrag}
\end{equation}

\noindent 
The ratio $v_\infty/R_\ast$ is a flow rate of
clumps over the scale of the stellar radius, which is the scale of the
wind density, whereas $\dot{N}_0$ is the injection rate of clumps into
the flow.  Hence the ratio $\dot{N}_0/(v_\infty/R_\ast)$ represents a relative
clustering of clumps along a radial.

In equation~(\ref{eq:nocovfrag}),
the integration over $\mu$ reflects the fact that the circular
shell fragments, when seen oblique to the normal, appear as ellipses
of uniform brightness $I_\nu$ (for a given fragment) and area
$\mu\, \delta \Omega \,R_\ast^2$.
Note that the optical
depth appears to diverge as $\mu \rightarrow 0$, which is unphysical;
however, in this model such edge-on fragments also have zero area
and so never contribute to the flux.  

The solution for
the flux would be analytic if the lower limit for the integration
in radius were to extend to zero.  As in previous works
\citep[c.f.,][]{1975MNRAS.170...41W}, changing the lower limit from 1 to 0
is acceptable for long wavelengths where $\tau_0 \gg 1$.
A general solution to integrals with the form of
equation~(\ref{eq:nocovfrag}) is 

\begin{equation}
\int_0^\infty \, \left[ 1 - e^{-ax^{-b}}\right] \, r^2\, dr
	= a^{3/b}\,\left[\frac{-\Gamma(-3/b)}{b}\right],
\end{equation}

\noindent if $b>3$.  
For the application of interest, the solution is

\begin{equation}
F_\nu = \pi\,B_\nu\,\frac{\delta \Omega\,R_\ast^2}{D^2}\,\frac{\dot{N}_0}
        {v_\infty/R_\ast} \,
	\int_0^1\, \left(\frac{\tau_0}{\mu}\right)^{3/2q}
	\,\left[\frac{-\Gamma(-3/2q)}{2q}\right]\,\mu\,d\mu.
\end{equation}

\noindent Evaluating the integral in $\mu$ and manipulating
the integration constants yields 

\begin{equation}
F_\nu = \frac{2q/3}{4q-3}\,\Gamma\left(1-\frac{3}{2q}\right)
	\, \tau_0^{3/2q} \times
	\pi\,B_\nu\,\frac{\delta \Omega\,R_\ast^2}{D^2}\,\frac{\dot{N}_0}
        {v_\infty/R_\ast}.
\end{equation}

\noindent Note that with $q=2$ for an inverse square law density,
the SED would be $F_\nu \propto g_\nu^{0.75}\,\lambda^{-0.5} \sim
\lambda^{-0.42}$.  This is a fairly shallow SED slope.  The well-known
SED slope of $-0.6$ for a smooth wind occurs when $q \approx 5/4$.

\subsection{Sparse Spherical Clumps}
\label{sec:sparsesph}


The results presented here for spherical clumps are an extension
of the method discussed in IC,
who considered an application of the free-free emission from
constant density, spherical clumps to explain anomalous SED slopes
seen in some ultra-compact H~II regions.  
Here the approach of IC is applied to a time-averaged spherical
wind.  Unlike the interstellar situation, application to a stellar wind
imposes a characteristic optical depth distribution for an ensemble of
clumps owing to the nature of the clumps partaking in a spherical outflow.
Consider the SED from a single isolated clump.  IC showed 
\citep[also in][]{1989agna.book.....O}
that the flux of emission from a single clump is given
by

\begin{equation}
F_\nu = \frac{R_\ast^2}{D^2}\,\pi\,B_\nu\,G(\tau_{\rm cl}),
	\label{eq:IC}
\end{equation}

\noindent where $\tau_{\rm cl}$ is the optical depth along the diameter of the
spherical clump which $\kappa_\nu\,\rho$ constant within its interior, as given
by 

\begin{equation}
\tau_{\rm cl} = 2\,\kappa_\nu(\rho_{\rm cl},\lambda)\,\rho_{\rm cl}\,R_{\rm cl},
\end{equation}

\noindent Note that equation~(\ref{eq:IC}) is based on equation~(4)
from IC, but corrects for an errant factor of 2 that should not
have been present in IC.

The function $G(\tau_{\rm cl})$ is given by

\begin{equation}
G(\tau_{\rm cl}) = 1 - \frac{2}{\tau_{\rm cl}^2}\,\left[ 1- (1+\tau_{\rm cl}) 
	\, e^{-\tau_{\rm cl}} \right].
	\label{eq:G}
\end{equation}

\noindent At large optical depth with $\tau_{\rm cl}\gg 1$, $G\approx 1$. 
At the opposite extreme of $\tau_{\rm cl} \ll 1$, $G\approx 2\tau_{\rm
cl}/3$, which is the area-averaged value across the sphere.  

The optical depth $\tau_{\rm cl}$ along the diameter of a clump located
at radius $\tilde{r}$ can be expressed as

\begin{equation}
\tau_{\rm cl}(\tilde{r}) = \tau_0(\lambda)\,\left(\frac{\rho_{\rm cl}^2\,R_{\rm cl}}
	{\rho_0^2\,R_0}\right),
\end{equation}

\noindent where $\rho_0$ and $R_0$ are fiducial values.  The density
within the clump is constant, but the question of how that density
varies with radius in the wind must be addressed.  One might consider
spherical clumps to maintain constant solid angle, like the fragment
clump case.  In order to maintain spherical shape,
this would imply that $R_{\rm cl} \propto r$, and so $\rho_{\rm cl}
\propto r^{-3}$.  However, the reference case 
involves a density that drops as $r^{-2}$.  Imposing $\rho_{\rm cl}
\propto r^{-2}$ means that spherical clumps do not 
expand at constant solid angle.

To allow for greater generality, the form $\rho_{\rm cl}
\propto \tilde{r}^{-m}$ is adopted.  Mass conservation for a clump requires
that $R_{\rm cl}^3 \rho_{\rm cl} = constant$, and so $R_{\rm cl}
\propto \tilde{r}^{m/3}$.  As a result, the position-dependent optical depth
of a spherical clump becomes

\begin{equation}
\tau_{\rm cl} = \tau_0(\lambda)\, \tilde{r}^{5m/3}.
\end{equation}

\noindent Note that spherical clumps have been used in considerations
by \cite{2007A&A...476.1331O} and \cite{2012ApJ...750...40I}.  It
is not uncommon for a spherical shell to ``fragment'' into elongated
features ala a Rayleigh-Taylor instability, and these in turn can
further break up into ball-shaped structures.  \cite{2005A&A...437..657D}
observe this effect in their 2D simulations for time-dependent
line-driven winds.  Their models show that wind structures develop
rings that are round in cross-section, rather than flat, lending
support to the idea that roughly spherical clumps could form.
Consequently, at any given time, a snapshot of the flow would possess
a component comprised of spherical clumps.

For the case of sparse clumps, the total emergent flux from the ensemble
is:

\begin{eqnarray}
F_\nu & = & \pi B_\nu\,\frac{R_{\rm cl}^2}{D^2}\,\int_1^\infty\,
	n_{\rm cl}(\tilde{r}')\,G(\tau_{\rm cl})\,4\pi\,\tilde{r}'^2 \, d\tilde{r}'\\
 & = & \pi B_\nu\,\frac{R_{\rm cl}^2}{D^2}\,\frac{\dot{N}_0}{v_\infty/R_\ast}
	\int_1^\infty\,G[\tau_{\rm cl}(\tilde{r}')]\,d\tilde{r}' .
\end{eqnarray}

At this point it is useful to adopt a change of variable from
normalized radius $x$ to optical depth $\tau_{\rm cl}$ for the
integration.  Let the largest value of $\tau_{\rm cl}$ be $\tau_{\rm
max}$ for a clump located at $x=1$.  Then $x=(\tau_{\rm cl}/\tau_{\rm
max})^{-3/5m}$. The integration that was over the clump ensemble
in spatial location becomes one over the clump ensemble in optical
depth space (similar in spirit to the work of IC), with

\begin{equation}
F_\nu = \frac{3}{5m}\,F_0(\lambda)\,\tau_{\rm max}^{\gamma-1}\,
	\int_0^{\tau_{\rm max}}\, \tau_{\rm cl}^{-\gamma}\,G(\tau_{\rm cl})\,
	d\tau_{\rm cl},
	\label{eq:nocover}
\end{equation}

\noindent where

\begin{equation}
\gamma = \frac{3+7m}{5m},
\end{equation}

\noindent and

\begin{equation}
F_0 = \pi B_\nu\,\frac{R_{\rm cl}^2}{D^2}.
	\label{eq:f0}
\end{equation}

One of the interesting attributes of equation~(\ref{eq:nocover})
is the wavelength dependence of the factors appearing in front of
the integral.  In the Rayleigh-Jeans limit, the product of the
Planck function and $\tau_{\rm max}$ yields a power law with

\begin{equation}
B_\nu\,\tau_{\rm max}^{\gamma-1} \propto g_\nu^{\gamma-1}\,\lambda^{-4+2\gamma}.
\end{equation}

\noindent The integral itself contributes an additional wavelength
dependence via the upper limit $\tau_{\rm max}$.  Given that $m=2$
and $m=3$ are the relevant density power-laws of interest, $\gamma$
equals 1.7 and 1.6, respectively.  The factors outside the integral,
using $g_\nu \propto \lambda^{0.11}$, give corresponding
wavelength dependences of $\lambda^{-0.53}$ for $m=2$ and
$\lambda^{-0.72}$ for $m=3$.  At long wavelengths such that $\tau_{\rm
max} \gg 1$, the integral factor has only a weak dependence on $\lambda$.
The reason is that the factor $\tau_{\rm cl}^{-\gamma}$ in the
integrand dominates such that most of the integration is set at low
and modest values of $\tau_{\rm cl}$.  At long wavelengths the
integral is well-approximated as a constant (but one that depends
on $m$), and the SED slope is determined primarily by the factors
outside the integral, which for $m=2$ and 3 bracket the canonical
result of $-0.6$ for microclumping.

Note that these results assume a universal size for clumps at
the inner wind radius.  It is straightforward to introduce a distribution
of clump sizes.  Suppose that ${\cal P}_{\rm cl}$ is the probability for
a clump to have a radius $R_{\rm cl}$ in the interval of $R_{\rm min}$
to $R_{\rm max}$.  Then the flux for an ensemble of sizes would proceed
as follows.  

The result of equation~(\ref{eq:nocover}) is now indicated by
$\hat{F}_\nu$ for the SED contribution from clumps that have
a size $R_{\rm cl}$.  For a range of sizes, the flux becomes

\begin{equation}
F_\nu = \int_{R_{\rm min}}^{R_{\rm max}}\,{\cal P}(R_{\rm cl})\,
	\hat{F}_\nu (R_{\rm cl})\, dR_{\rm cl},
\end{equation}

\noindent where the probability distribution is normalized, with

\begin{equation}
\int_{R_{\rm min}}^{R_{\rm max}}\,{\cal P}(R_{\rm cl})\,dR_{\rm cl} = 1.
\end{equation}

The influence of a range of sizes affects the SED calculation in two
ways.  First, there is an overall factor of $R_{\rm cl}^2$ in $F_0$, so
smaller clumps tend to be less bright.  Second, assuming that all clumps
have the same initial density, smaller clumps tend to be more optically
thin, which makes them less bright, and further means that they
become optically thick at longer wavelengths as compared to bigger clumps.

\begin{figure}
\includegraphics[width=\columnwidth]{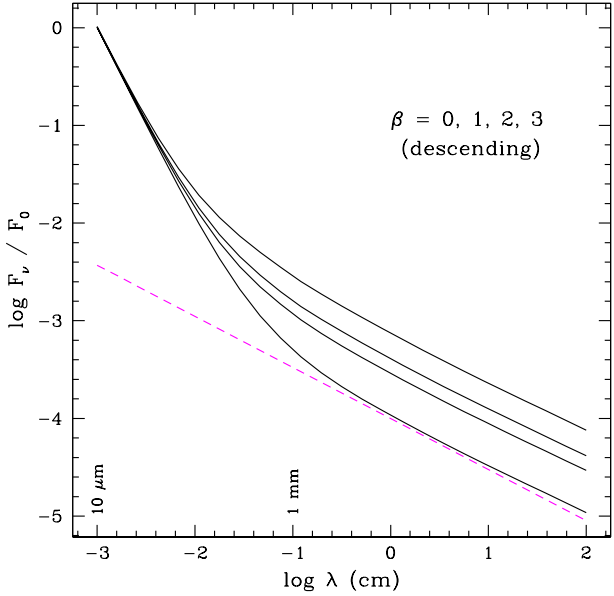}
\caption{SEDs for spherical clumps in the sparse limit with a distribution in
clump sizes $R_{\rm cl}$.  In these four models, $\beta$ signifies 
power-law distributions for clump sizes.  The
case of $\beta=0$ is a flat distribution (clumps of different sizes
are all equally likely, within a specified range); increasing
$\beta$ results in increasingly more small clumps as compared
to large ones.  All the
models have the same number of clumps and same optical depth parameter.
\label{fig2}}
\end{figure}

Figure~\ref{fig2} shows SEDs for a clump distribution, ${\cal P}_{\rm
cl} \propto R_{\rm cl}^{-\beta}$, with values of $\beta= 0, 1, 2,$
and 3, from top to bottom.  The examples are designed to display
the relative effect from having a greater representation of small
clumps.  The magenta indicates the asymptotic SED value.  These
models are for $m=2$.  An ensemble of clump sizes does not change
the SED asymptote, only the brightness level and where in wavelength
the wind emission begins to dominate the stellar component.

\section{Winds with Macroclumping:  Porosity Approach}
\label{sec:macro}

The preceding discussion of the sparse limit 
shows that the geometry of the clumps can affect
both the SED slope and the flux level.  However, one generally
expects that multiple clumps will lie along a given sightline.  In such
cases of overlap, the radiative transfer is more complicated.  The
emission by each rearward clump along a los must be attenuated by
all the clumps intervening between it and the observer, with 
corresponding emission increments for each.

Several researchers have considered the radiative transfer in porous,
massive star winds both for diagnostics such as X-ray and UV lines
\citep{2003A&A...403..217F,2004A&A...422..675O,2006ApJ...648..565O,
2007A&A...476.1331O,2012MNRAS.420.1553S,2014A&A...568A..59S},
and in terms of wind-driving physics such as super-Eddington and
line-driven flows \citep{2004ApJ...616..525O,2014A&A...568A..59S}.
Here the use of the porosity formalism is brought to bear on the
problem of continuum free-free emission.

The porosity approach is a way of expressing discrete sums over
clump structures for the radiative transfer along a sightline in
terms of integral expressions.  The result reduces to the microclumping
case when clumps are optically thin.  It is when individual clumps
become optically thick that the clump geometry must be considered.
The ability to move from discrete sums to integral expressions is
the opposite of the sparse limit: for the sparse limit, the radiative
transfer is confined to individual clumps; porosity allows for
multiple clumps along a sightline.

\cite{2003A&A...403..217F} showed that the effective optical depth
along a sightline through a porous wind is given by

\begin{equation}
\tau = \int \, \left[ \int n_{\rm cl}\,R_\ast^3\,
	\left(1-e^{-\tau_{\rm cl}}\right)\,d\tilde{A}_{\rm cl} \right]\,
	d\tilde{z},
	\label{eq:tclumps}
\end{equation}

\noindent where $\tilde{A}_{\rm cl}$ is the normalized projected
area of the clump toward the observer.  The inner integration is
across the face of the clump; the outer is through the wind.  The
interpretation is that the product $n_{\rm cl}\, R_\ast^3\,d
\tilde{A}_{\rm cl}d\tilde{z}$ is the number of clumps encountered
in a differential unit of volume.  The factor $(1-\exp(-\tau_{\rm
cl}))$ is the fraction of light absorbed by a clump at this location,
with $\tau_{\rm cl}$ the optical depth of the clump along the ray.
Clearly, if $\tau_{\rm cl} \ll 1$, the effective optical depth is just
the cumulative optical depth for the los in the interval $dz$.
However, when $\tau_{\rm cl} \gg 1$, the effective optical depth
by clumps.

Regarding this last point, the normalized porosity length is defined
as $\tilde{h}=h/R_\ast = (n_{\rm cl}\, A_{\rm cl}\,R_\ast)^{-1}$
and represents the mean free path between clumps along a sightline.
For thick clumps, the effective optical depth becomes

\begin{equation}
\tau = \int \, d\tilde{z}/\tilde{h}.
\end{equation}

\noindent This amounts to the number of mean free paths along the los.
Consequently, even if clumps are quite optically thick, it may be that
$\tau$ can be relatively small if the environment is highly porous (i.e.,
$\tilde{h}\gg 1$).

\subsection{Porosity with Fragments}

\begin{figure*}
\includegraphics[width=\columnwidth]{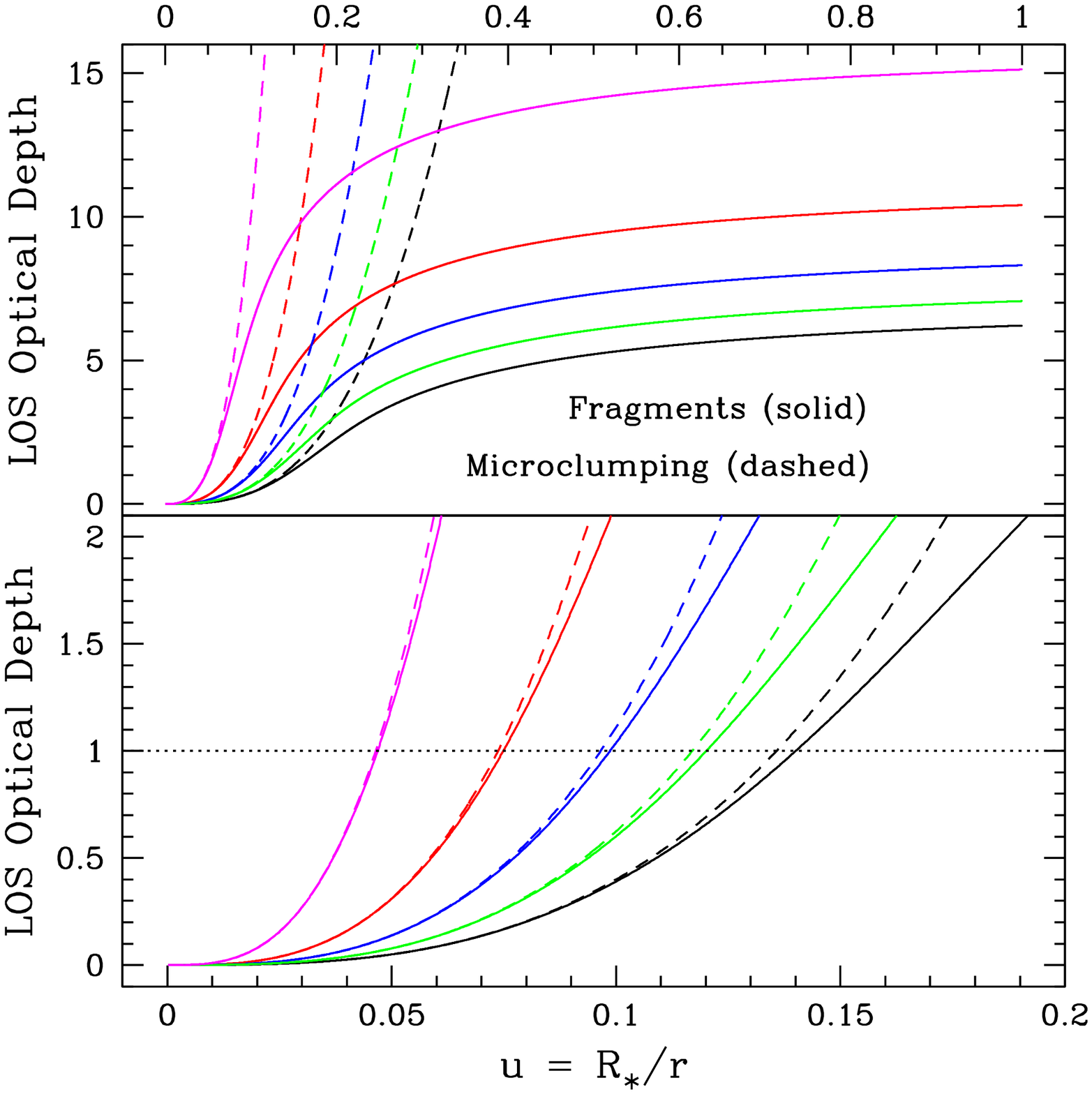}
\includegraphics[width=\columnwidth]{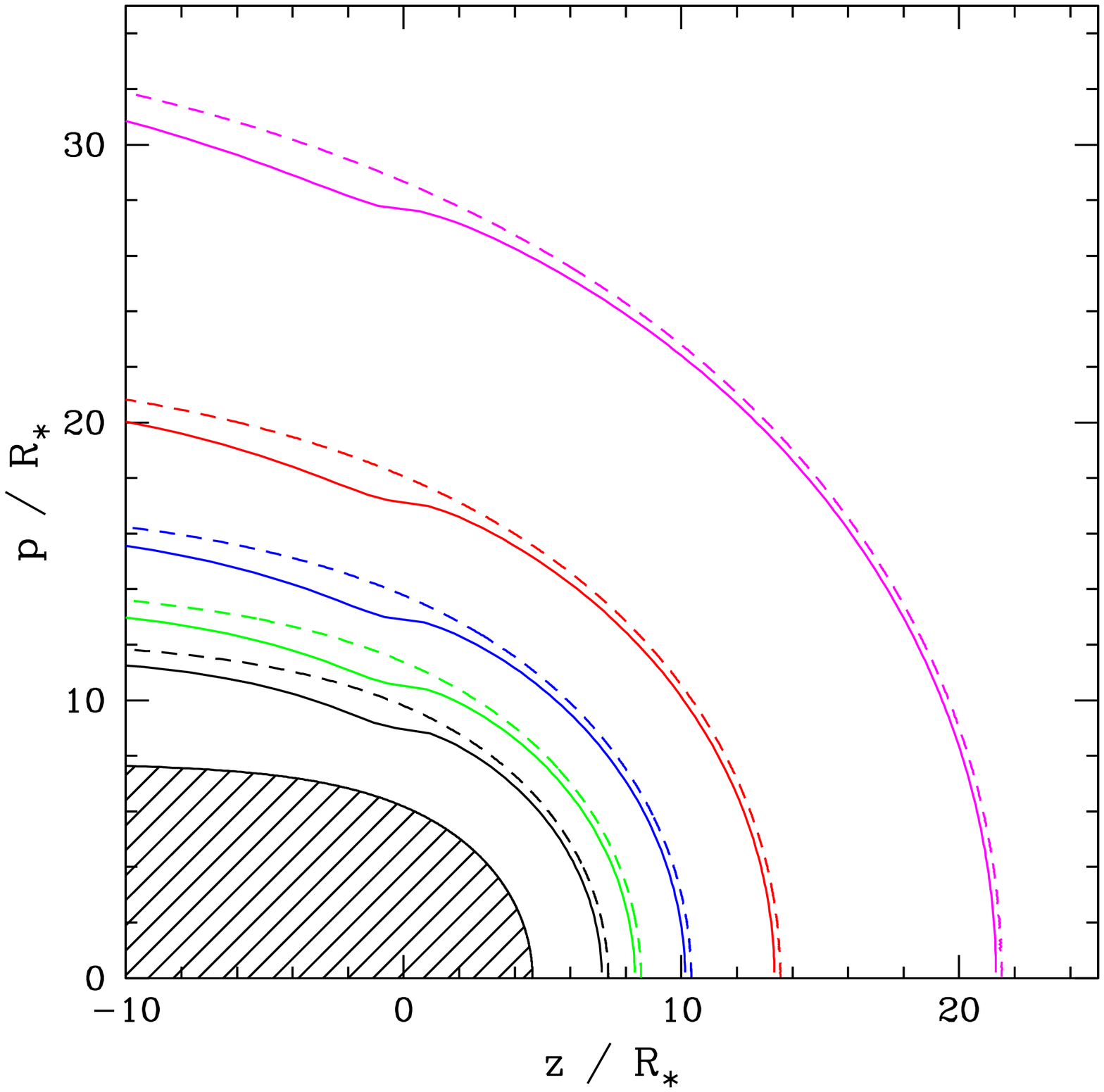}
\caption{
Left:  The line-of-sight (los) optical depth as a function
of normalized inverse radius $u$ as compared with microclumping
for a range of volume filling factors.  
Upper panel is for the full range of $u$ from 0 to 1; lower is
for small $u$ (i.e., $r\gg R_\ast$) to highlight the region where
microclumping and porosity diverge.  The volume filling factors
are for $f_0 = 0.01$ (magenta), 0.04, 0.09, 0.16, and 0.25 (black).
Right:  Associated contours for optical depth unity in the $z-p$ plane.
The color scheme and line types are the same as in the left panel.
\label{fig3}}
\end{figure*}

\begin{figure}
\includegraphics[width=\columnwidth]{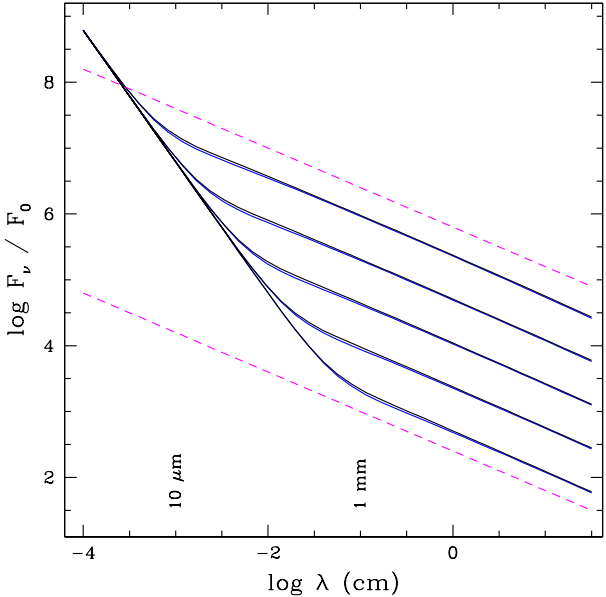}
\caption{
SED calculations that compare microclumping models with porosity
models for shell fragments.  All curves are for $\tau_0=300$, with
$f_0 = 0.0001$ (top), 0.001, 0.01, 0.1, and 1.0 (bottom).
The two sets of curves are so close as to be indistinguishable in
this plot.
}
\label{fig4}
\end{figure}

Consider again the pancake-shaped shell fragment in projection.
For a geometrically thin clump of not overly large solid angle,
such a fragment appears elliptical in shape and is uniformly bright
across its projected face.  The porosity length for a distribution
of fragments is

\begin{equation}
\tilde{h} = \frac{1}{n_{\rm cl}\,\mu\,A_{\rm cl}\,R_\ast} =
	\left(\frac{v_\infty/R_\ast}{\dot{N}_0}\right)\,
	\left(\frac{4\pi}{\mu\,\delta \Omega}\right).
\end{equation}

\noindent The shell fragments have a porosity length that depends
on $\mu$, or location around the star.  For the outer wind that is
at terminal speed, $\tilde{h}$ does not vary with distance from the
star, but does vary with location about the star. The associated
volume filling factor for shell fragments is

\begin{equation}
f_V = n_{\rm c}\,V_{\rm c} = n_{\rm c}\,A_{\rm c}\,R_\ast
	\,\delta \tilde{l}_0 = \frac{\dot{N}_0}{v_\infty/R_\ast}\,
	\left(\frac{\mu\,\delta \Omega}
	{4\pi}\right)\,\delta \tilde{l}_0,
\end{equation}

\noindent where $V_{\rm c}$ is a volume over which the filling
factor is determined, and $\delta \tilde{l}_0$ is the normalized radial
width of the fragment.  This width could be a function of radius,
but will be taken as constant in the discussion that follows.  

The expression for the optical depth to position $u=1/x$ along
the los to the star center (i.e., $\mu=1$) is given by

\begin{equation}
\tau = \frac{\tau_0}{f_0}\,\int_0^u \left[\frac{1-e^{-\tau_{\rm cl}}]}
	{\tau_{\rm cl}}\right]\, u^2\,du,
	\label{eq:tfrag}
\end{equation}

\noindent where $f_V = f_0\,\mu$, 

\begin{equation}
f_0 = \frac{\dot{N}_0}{v_\infty/R_\ast}
        \left(\frac{\delta \Omega}{4\pi}\right)\,\delta \tilde{l}_0,
\end{equation}

\noindent and

\begin{equation}
\tau_{\rm cl} = \left(\frac{\tau_0}{f_0}\right)\,
	\left(\frac{\delta \tilde{l}_0}{f_0}\right)\,\left(\frac{u^4}{\mu}\right).
\end{equation}

\noindent The bracketed factor in equation~(\ref{eq:tfrag}) 
is the form of an escape probability.  At low
values of $\tau_{\rm cl}$, the factor reduces to unity, in which
case the microclumping limit is recovered.  For large $\tau_{\rm
cl}$, the factor reduces to $1/\tau_{\rm cl}$,  
so that the effective optical depth when clumps become thick
can be much suppressed with macroclumping as compared to micrcolumping.

\begin{figure*}
\includegraphics[width=\columnwidth]{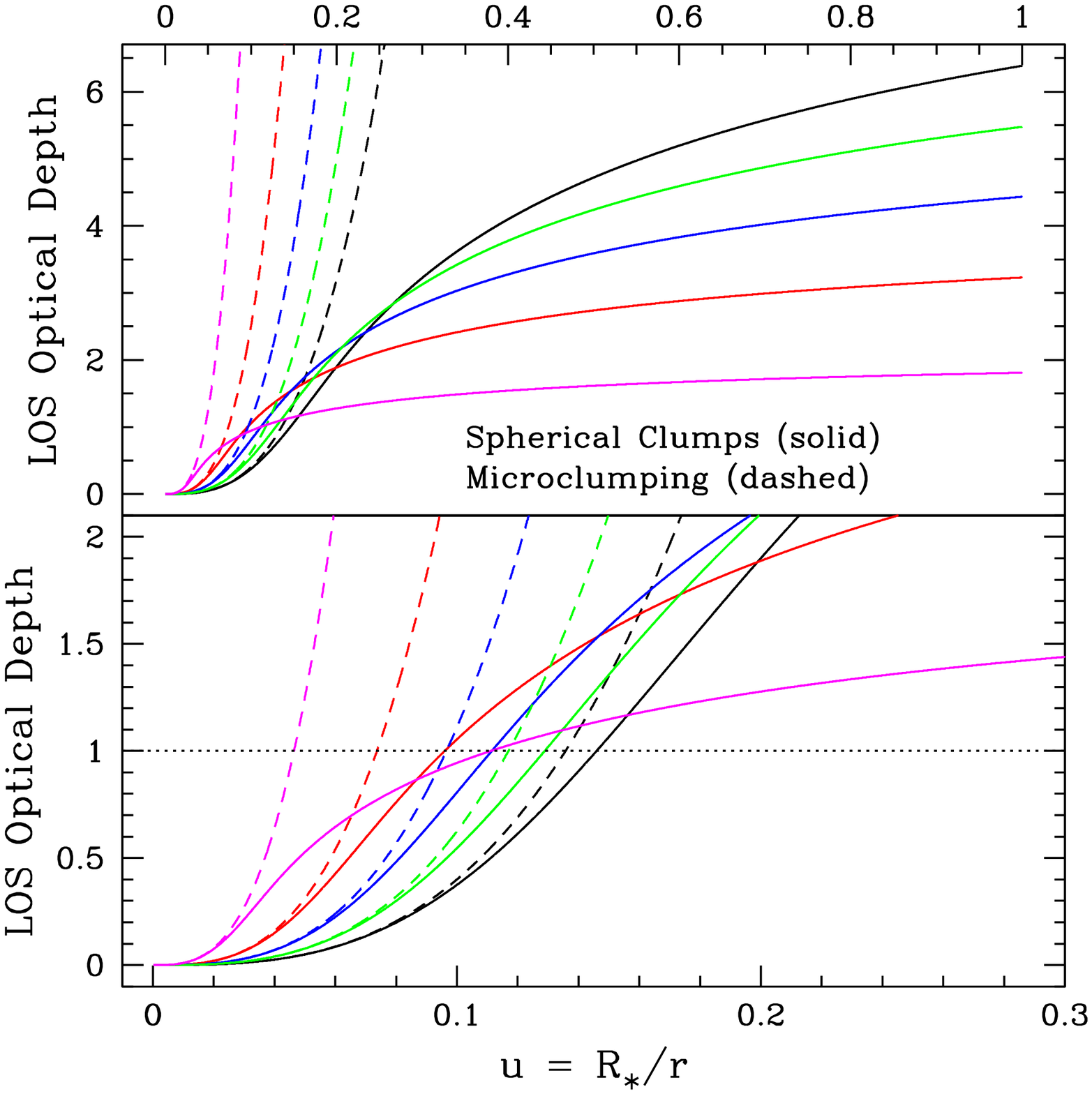}
\includegraphics[width=\columnwidth]{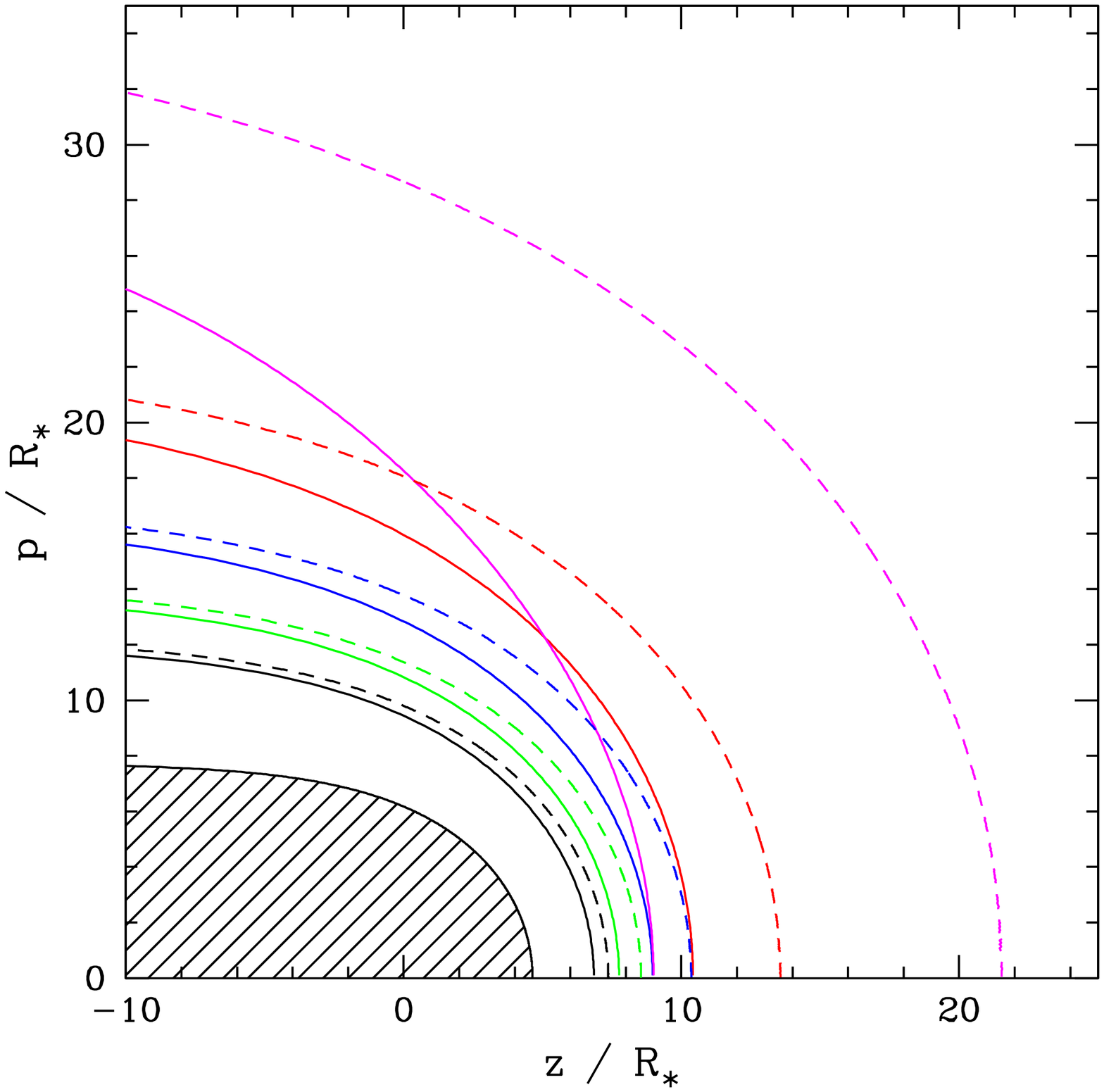}
\caption{
The models here are like those in Fig.~\ref{fig3}, but for porosity
with spherical clumps.  The models have the same value of $\tau_0=300$,
$f_V$ ahs the same range of values as did $f_0$, with the same
corresponding color designations.  Major differences as compared
to shell fragments are (a) the overall lower optical depths achieved at
the star, (b) the fact that departures between microclumping and the
porosity start at lower optical depths of only a few tenths, and
(c) the crossing of curves, both in the los optical depths (left) and
in some cases the contour curves (right).
\label{fig5}}
\end{figure*}

The left side of Figure~\ref{fig3} shows the optical depth for a
sightline along a radial to the star based on equation~(\ref{eq:tfrag}).
The figure compares the porosity case against the microclumping
assumption, ignoring the fact that the clumps are optically thick.
The volume filling factors are $f_0 = 0.01$ (magenta), 0.04, 0.09,
0.16, and 0.25 (black), all with $\tau_0 = 300$.  Solid lines are
for the shell fragments; dashed lines are for microclumping using
the same parameters. The radial width of a shell fragment is chosen to
be $\delta \tilde{l}_0= f_0$.  The upper panel shows
curves for the full range of $u$; lower is a blow-up around the
region of optical depth unity.  Differences between microclumping
and porosity with shell fragments only begins to develop near optical
unity; departures between the two then increase toward higher optical
depth.  

The right side of Figure~\ref{fig3} displays a cross-section of the
axisymmetric optical-depth unity contours in the $z-p$ plane.  The
colors and line types correspond to those of the left side of the
figure.  The central hashed region is for a smooth, unclumped wind
with the same value of $\tau_0$.  Note that the contours are fairly
closely matched, save for the ``dimple'' that results for the shell
fragments.  That feature arises where shell fragments are seen more
nearly edge-on so that the porosity length is large.

Figure~\ref{fig4} shows model SEDs for both microclumping and for
porosity with shell fragments.  The optical depth parameter
$\tau_0=300$ is the same as in Figure~\ref{fig3}, but a greater
range of volume filling factors are used to spread out the transition
in wavelength from star-dominanted to wind-dominated continua.  For
Figure~\ref{fig4}, the filling factor constant ranges from $f_0=10^0$
(lowest curve) to $10^{-4}$ (highest curve).  Note that the transition
from star-dominated to wind-dominated would sample the inner,
accelerating portion of the wind, which is not currently included
the model.  Such effects are ignored to emphasize comparisons between
microclumping and macroclumping effects.

The two sets of SEDs are so close as to be indistinguishable in the
figure.  That microclumping and porosity with shell fragments yield
essentially identical SEDs aligns well with the fact that the two
models have such close los optical depths for the optically thin
portion of the wind.  

\subsection{Porosity with Spherical Clumps}

The nature of spherical clumps is different from the case of shell
fragments.  For spheres the volume filling factor is given by

\begin{equation}
f_V = n_{\rm cl}(r)\,V_{\rm cl} = \frac{1}{3}\,\frac{\dot{N}_0}
	{v_\infty/R_\ast}\,\tilde{R}_0^3,
\end{equation}

\noindent where $n_{\rm cl} \propto r^{-2}$, $V_{\rm cl} \propto
r^2$, $\tilde{R}_0$ is the normalized radius of a clump
as the base of the wind. 
Whereas shell fragments have a volume filling factor that
does not depends on radius from the star, but does
depend on location about the star, owing to their flattened shapes,
the $f_V$ for spheres is the same at all locations.
The related porosity length scales as:

\begin{equation}
\tilde{h} = \frac{1}{n_{\rm cl}\,A_{\rm cl}\,R_\ast} =
	4\,\left(\frac{v_\infty/R_\ast}{\dot{N}}\right)\,
	\left(\frac{1}{\tilde{R}_0^2}\right)\,\tilde{r}^{2/3} = \frac{4}{3}\,
	\frac{\tilde{R}_0}{f_0}\,u^{-2/3}.
\end{equation}

\noindent which has the inverse scaling with radius as the clump
solid angle.  Consequently, the porosity length becomes larger
with increasing distance from the star.

For spheres the lines of sight intersect the spherical clumps along
different chords.  Noting that $\tau_{\rm cl}$ is the optical depth
along the diameter of a spherical clump, Appendix~\ref{appB} details
the analytic derivation for the inner integral for
equation~(\ref{eq:tclumps}).  The result, simply stated here, is
that

\begin{equation}
\tau = \int \, n_{\rm cl}\, A_{\rm cl}\,R_\ast\,G(\tau_{\rm c})
	\,dz,
\end{equation}

\noindent where 

\begin{equation}
G(\tau_{\rm cl}) = 1-\frac{2}{\tau_{\rm cl}}\,\left[1-e^{-\tau_{\rm cl}}
	- \tau_{\rm cl}\,e^{-\tau_{\rm cl}}\right].
\end{equation}

\noindent The optical depth integral can be more conveniently
expressed as

\begin{equation}
\tau = \frac{\tau_0}{f_V}\,\frac{1}{\tilde{p}^3}\,\int_0^\theta\,
	\left[\frac{3G(\tau_{\rm cl})}{2\tau_{\rm cl}}\right]
	\sin^2\theta\,d\theta,
	\label{eq:tau_theta}
\end{equation}

\noindent where as before, $\tan \theta = \tilde{p}/\tilde{z}$.
The ratio $3G/2\tau_{\rm c}$ acts like an escape probability for
the clump.  When $\tau_{\rm cl} \ll 1$, the ratio $3G/2\tau_{\rm
cl}$ reduces to unity, as expected for optically thin clumps.  When
$\tau_{\rm cl}\gg 1$, the ratio becomes $1/(2\tau_{\rm c}/3)$.  The
factor of $2/3$ arises from area-averaging the optical depth across
the face of the spherical clump.

\begin{figure}
\includegraphics[width=\columnwidth]{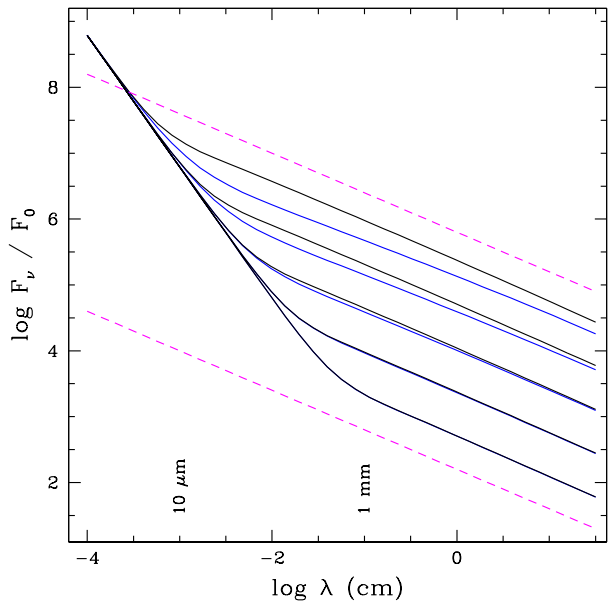}
\caption{
Model SEDs to compare microclumping (black) and porosity with
spherical clumps (blue).  The models are for $\tau_0$ and $f_V$
values used in Fig.~\ref{fig5}.  The SEDs with spherical clumps
achieve the same asymptotic slopes as for microclumping, but at
lower flux levels.  The discrepancy in the flux level worsens as
$f_V$ becomes smaller.
}
\label{fig6}
\end{figure}

Along the los to the star, the optical depth expression becomes

\begin{equation}
\tau(u) = \frac{\tau_0}{f_V}\,\int_0^u \,\left[\frac{3G(\tau_{\rm cl})}
	{2\tau_{\rm cl}}\right]\,u^2\,du,
\end{equation}

\noindent The optical depth across the diameter of the clump is

\begin{equation}
\tau_{\rm cl} = \left(\frac{\tau_0}{f_V}\right)\,
	\left(\frac{2\tilde{R}_{\rm cl}}{f_V}\right)\,u^{10/3}.
\end{equation}

\noindent The unusual power-law exponent with radius derives from
how the spherical clump evolves in size throughout the wind, as
explained next.

To illustrate the effects of porosity with spherical clumps,
Figure~\ref{fig5} shows the los optical depth in the porosity
formalism.  The figure is similar to Figure~\ref{fig3} for shell
fragments, with the same $\tau_0$, the same range in $f_V$, and the
same color scheme.  Again, the dotted curves are for microclumping.
The clump radii scale as $\tilde{R}_0 = 0.03 \, (f_V/0.01)^{1/3}$.
For low optical depths, the curves closely follow each other as
expected.  Departures between microclumping and porosity develop
around los optical depth of a couple tenths.  When $f_V$ is small,
optical depth unity for the spherical clumps occurs at a notably
smaller radius than for microclumping; for shell fragments optical
depth unity occured close in radius to the microclumping result.

The right side of Figure~\ref{fig5} displays corresponding optical
depth unity contours for the $z-p$ plane.  The colors are the same
as for the left figure.  Rearward of the star (large negative values of
$z$), the contours are sequenced in terms of $f_V$, with the highest
filling factor (least porous wind) being more compact, and smallest
filling factor (most porous wind) being larger.  The optical
depth unity contours for microclumping with the same value of
$\tau_0$ and $f_V$ can lie at larger radii when $f_V$ is
small.  Consequently, porosity with spherical clumps indicates
that the bulk of the long-wavelength emission arises from a more
compact region of the wind as compared with microclumping.

Figure~\ref{fig6} shows SED calculations that contrast microclumping
with a porous wind consisting of spherical clumps.  The microclumping
results are the black lines, and the porous wind results are shown
in blue.  Unlike the case for shell fragments, there can be noticeable
differences in flux levels.  There are several key points to
be made about these results.

\begin{figure}
\includegraphics[width=\columnwidth]{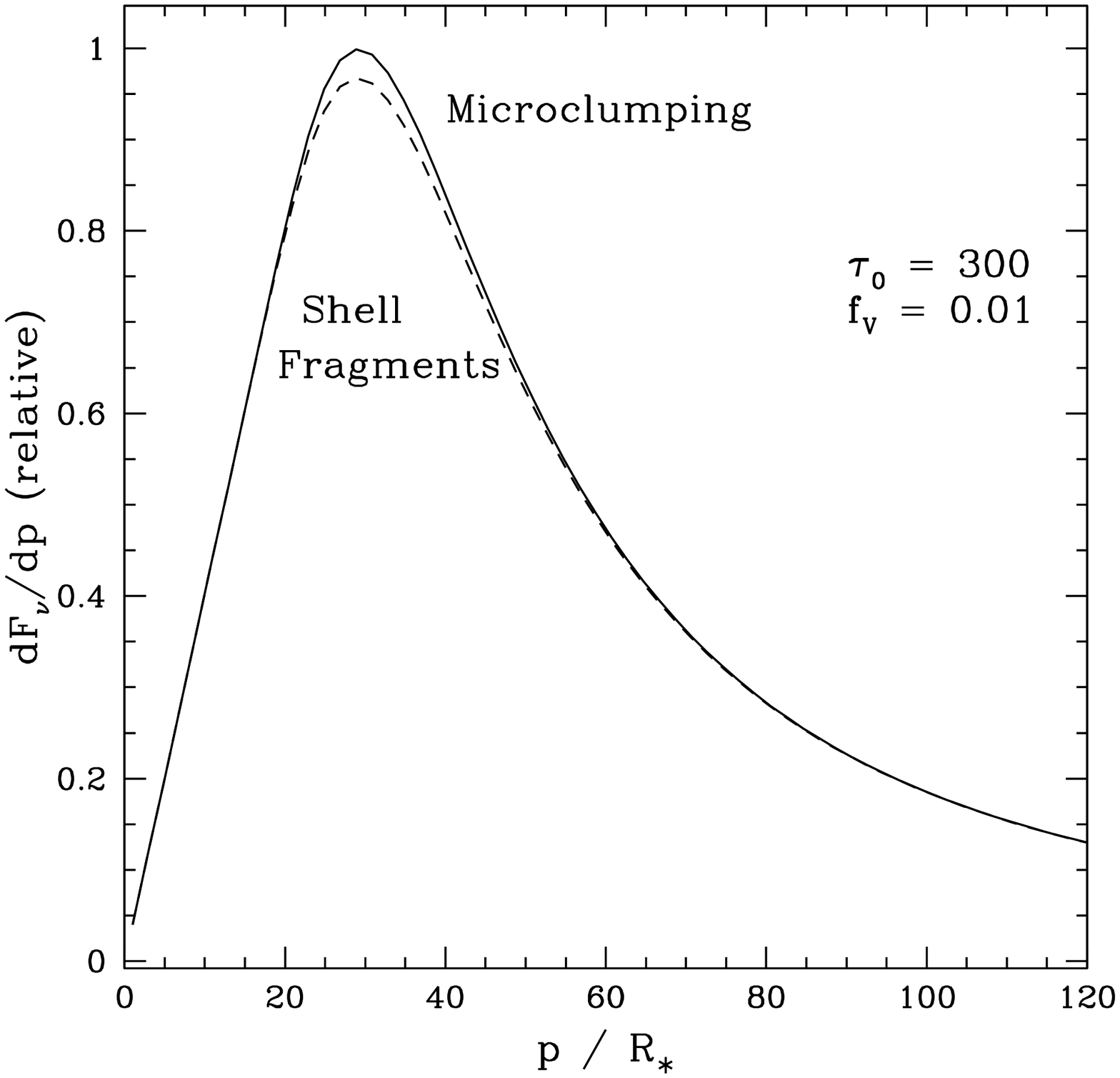}
\includegraphics[width=\columnwidth]{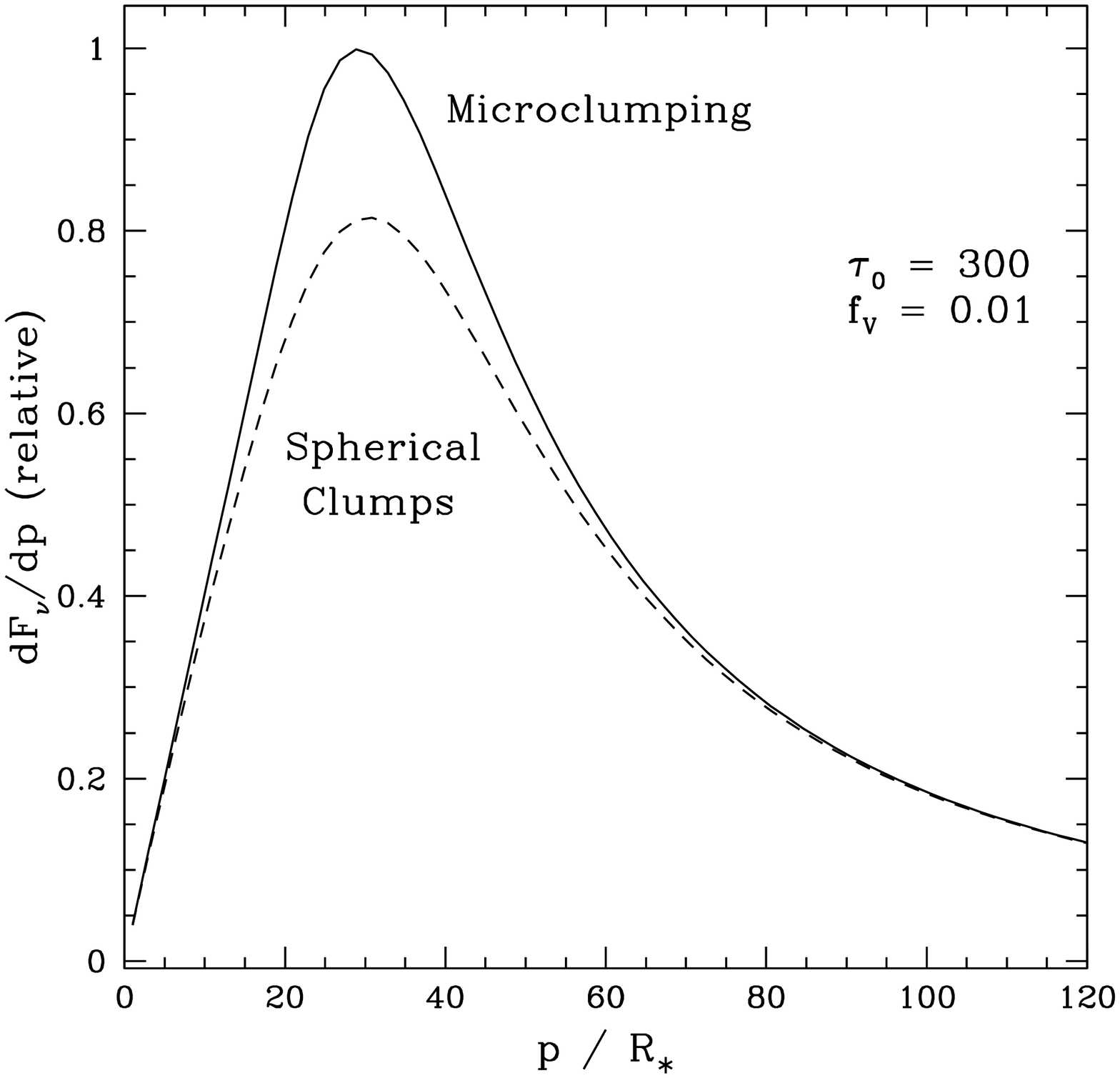}
\caption{A comparison of SEDs with porosity for shell fragments (upper)
versus spherical clumps (lower) in terms of 
$dF_\nu/dp$ (see text) versus impact parameter $p/R_\ast$.
For both panels the solid curve is for
microlensing, and the dashed curves include porosity effects.  The models
are for $\tau_0 = 300$ and $f_V = 0.01$.  The curve for shell fragments
much more closely follows the curve for microclumping; the
curve for spherical clumps shows a greater deviation.  The smaller
fluxes with porosity arise from the smaller effective optical
depths as compared to microclumping.  At large $p$, the curves
closely match where clumps become optically thin; at small
$p$, the curves closely match because even with porosity, the
effective optical depth becomes large.
\label{fig7}}
\end{figure}

\begin{itemize}

\item[--] There is no distinction between microclumping and porosity
when $f_V=1$ (lowest curve).

\item[--] When departures between microclumping and porosity do arise,
the differences are greater for smaller values of $f_V$.

\item[--] The differences are greatest at wavelengths around the
transition from star-dominated to wind-dominated SEDs.

\item[--] The SEDs with porosity tend toward the 
SEDs with microclumping at sufficiently long wavelengths.

\item[--] At wavelengths where departures are significant, the SED slope
is {\em more shallow} than the canonical slope of $-0.6$ value for
the power-law SED.

\item[--] Figure~\ref{fig7} contrasts the
cases of shell fragments and spherical clumps.  Plotted is the
relative differential flux contribution $dF_\nu/dp$ with impact
parameter $p$ as normalized by $R_\ast$, with

\[ \frac{dF_\nu}{dp} \propto p\, \left[1-e^{-\tau(p)}\right] . \]

\noindent The upper panel compares shell fragments (dashed) with
microclumping (solid); lower compares spherical clumps (dashed)
with microclumping (again solid).  The curve for shell fragments
closely matches the one for microclumping; with spherical clumps
there is a greater difference.  The latter arises because the run
of total effective optical depth achieved with spherical clumps is
lower than for shell fragments.  

\end{itemize}

In Figure~\ref{fig6}, with $f_V = 0.0001$, the slope for the porous
model SED, between $\log \lambda$ of $-1$ and $+1$, is very close
to $-0.55$, which is the value expected in the sparse limit for an
inverse square density (i.e., $m=2$) from \S~\ref{sec:sparsesph}.
That departures between the microclumping and porosity SEDs increase
with decreasing $f_V$ is a reflection of the idea that the sparse
limit is becoming a better approximation for the clump distribution.
At sufficiently long wavelengths, there are more optically thick
clumps, and the sparse limit no longer remains valid.

For a given value of $\dot{M}$, a porous wind with spherical clumps
can yield a lower flux level as compared to microclumping.  Ignoring
porosity effects and adopting the assumption of microclumping would
then require a lower value of $\dot{M}$ (or, a larger value of
$f_V$) to match the observed SED flux level.  However, the effect
is only significant at fairly extreme values of $f_V$.  With $f_V
= 0.001$, the SED with porosity differs from the microclumping one
by less than 0.2 dex.  Also, the severity of the departure between
the cases depends on $\lambda$.

\section{Conclusion}
\label{sec:conc}

All else being equal (isothermal, constant velocity, no ionization
gradients), the geometry of clumps can influence IR/radio fluxes
and power-law SED slopes when porosity effects are included.  One
should generally expect macroclumping effects to become important at
sufficiently long wavelengths, since the free-free opacity grows as
$\lambda^2$.

If the clumps take the form of shell fragments, both the SED flux
levels and SED slope are essentially identical to the microclumping
case.  For the shell fragment geometry, the los optical depth with
porosity close tracks with that of the microclumping case below
optical unity.  The optical unity contours closely follow those for
microclumping.  In the ``pseudo-photosphere'' picture, the optical
depth unity surface for microclumping and for porosity with shell
fragments are very similar in spatial extent, and the resultant
fluxes end up nearly the same over a wide range of volume filling
factors.

By contrast, spherical clumps can yield results that are distinct
from the case of microclumping.  Unlike the case of shell fragments
that have a constant porosity length in the asymptotic wind,
$\tilde{h}$ is not constant for spherical clumps for an inverse
square law density.  The case of porosity with spherical clumps
still achieves the canonical slope of $-0.6$, but departures between
SEDs with porosity and those that assume microclumping can arise
at some wavelengths.  The range of wavelengths over which the
differences occur, and the amplitude of the difference depends
strongly on the value of $f_V$.  Achieving a flux difference in
excess of 0.1~dex requires $f_V < 0.01$; significant differences
only result for rather extreme volume filling factors, of order
$10^{-4}$.  And, when those differences
arise, it is because the so-called ``sparse limit'' is being achieved.
The result is a power-law SED slope that, for an inverse square law
density, is more shallow than the canonical value of $-0.6$.



\section*{Acknowledgements}

Ignace is grateful to an anonymous referee for several helpful
comments, and to Maurice Leutenegger for having made a number of
insightful points on an earlier manuscript.  Ignace recognizes
support for this research through a grant from the National Science
Foundation (AST-0807664).

\appendix

\section{Analytic SED Solutions for Smooth Wind Cases}
\label{appA}

In the limiting case of large optical depth, \cite{1975MNRAS.170...41W}
and \cite{1975A&A....39....1P}) showed that the integration for the
flux of wind emission becomes analytic when $\kappa_\nu\rho \propto
r^{-4}$, for $r$ the radius in the wind.  In this case the optical
depth along a los becomes $\tau_\nu = a_0\,\tau_0(\lambda)\,
\tilde{p}^{-3}$, where $a_0$ is a constant, $\tau_0$ is the optical
scaling at wavelength $\lambda$, and $\tilde{p}$ is the normalized
impact parameter for the sightline.  The approach can be generalized
for $\tilde{p}$ raised to other powers.

The integral to be evaluated is:

\begin{equation}
F_\nu = 2\pi\,\frac{R_\ast^2}{D^2}\,B_\nu(T_{\rm w})\,
	\int_0^\infty\,\left(1-e^{-k_\lambda \,\tilde{p}^\alpha}\right)\,
	\tilde{p}\,d\tilde{p},
\end{equation}

\noindent where $k_\lambda \propto \tau_0(\lambda)$ and $\alpha$
is the power-law exponent.

This integral has an analytic
solution of

\begin{equation}
F_\nu = 2\pi\,\frac{R_\ast^2}{D^2}\,B_\nu(T_{\rm w})\,\left[
	-\frac{1}{\alpha}\,\Gamma\left(\frac{-2}{\alpha}\right)\,
	k^{-2/\alpha}\right]
\end{equation}

\noindent If $\alpha >2$, the recursion relation $\Gamma(1+x) = x\Gamma(x)$
can be used to show that the bracketed term is indeed positive, with

\begin{equation}
-\frac{1}{\alpha}\,\Gamma\left(\frac{-2}{\alpha}\right) = \frac{2}{\alpha^2}
	\, \Gamma\left(\frac{\alpha-2}{\alpha}\right).
\end{equation}

\section{The Spherical Macroclump}
\label{appB}

Discussions of macroclumping in stellar winds often adopt isotropic
or fragment clumps.  An analytic solution can be derived
for a spherical clump of uniform density that takes account of
variations in optical depth with chord length for use in the porosity
formalism.  The optical depth along a chord through a spherical
clump is

\begin{equation}
t_{\rm cl} = \frac{\tau_0}{f_V^2}\times 2\,\delta \tilde{z}\,\times \tilde{r}^{-4},
\end{equation}

\noindent where all lengths are scaled to the stellar radius $R_\ast$,
and $2\delta \tilde{z}(\tilde{\varpi})$ is the pathlength for a
chord at impact parameter, $\tilde{\varpi}$.  Using eq.~(\ref{eq:tclumps}),
the optical depth through a wind consisting of spherical clumps
is

\begin{equation}
\tau = \int n_{\rm cl}(r)\,R_\ast^3\, \left\{
	\int_0^{\tilde{R}_{\rm cl}} \left[ 1 - e^{-t_{\rm cl}(\delta \tilde{z})} \right]
	\,2\pi\,\tilde{\varpi}\,d\tilde{\varpi}\,\right\}\,d\tilde{z},
\end{equation}

\noindent for $n_{\rm cl}$ the number density of clumps.  The inner
integral relates to the area-weighted probability of a photon being
absorbed by the clump with optical depth $t_{\rm cl}$.  Noting that
$\delta z = \sqrt{\delta x_{\rm c}^2-\varpi^2}$, the preceding
integral can be written as

\begin{equation}
\tau = \int n_{\rm cl}(\tilde{r})\,R_\ast^3\, 
        \left[\pi\,\tilde{R}_{\rm cl}^2 - 2 \pi\,\int_0^{\tilde{R}_{\rm cl}} e^{-q(\tilde{r})\,\delta \tilde{z}}
        \,\delta \tilde{z}\,d\delta \tilde{z}\right]\,d\tilde{z},
\end{equation}

\noindent where a change of variable has been used to eliminate $\tilde{\varpi}$
in favor of $\delta \tilde{z}$.  
The second term in brackets is of the form

\begin{equation}
\tau = \int\,x\,e^{-qx}\,dx.
\end{equation}

\noindent This expression has an analytic solution, after
which the los optical depth reduces to

\begin{equation}
\tau = \int_{\tilde{z}_0}^\infty n_{\rm cl}(x)\,R_\ast^3\, \, \left\{ 1 -
	\frac{2}{\tau^2_{\rm cl}}\,\left[1-(1+\tau_{\rm cl})
	\,e^{-\tau_{\rm cl}}\right]\right\}\, d\tilde{z},
\end{equation}

\noindent with $\tau_{\rm cl}$ the optical depth across the diameter of a clump
located at radius $r$.  The term in braces is $G(\tau_{\rm cl})$
from eq.~(\ref{eq:G}).
The final result for the los optical depth is

\begin{equation}
\tau = \int_{\tilde{z}_0}^\infty \left[\frac{2G(\tau_{\rm cl})}{3\tau_{\rm cl}}
	\right]\,u^4\,d\tilde{z},
\end{equation}

\noindent where $u=\tilde{r}^{-1}$.  A change of variable from
$\tilde{z}$ to polar angle $\theta$ yields eq.~(\ref{eq:tau_theta}).

\bibliographystyle{mnras}
\bibliography{revised_fnl}

\end{document}